# First-principles modeling of electrostatics and transport in 2D topological transistors

Hyeonseok Choi[1], Yosep Park[2], Subeen Lim[1], and Yeonghun Lee[1,2,3*]

[1] Department of Electronics Engineering, Incheon National University, Incheon 22012, Republic of Korea

[2] Department of Intelligent Semiconductor Engineering, Incheon National University, Incheon 22012, Republic of Korea.

[3] Research Institute for Engineering and Technology, Incheon National University, Incheon 22012, Republic of Korea.

**Corresponding author**

[*] Email: y.lee@inu.ac.kr

**Abstract**

We develop a simulation framework for electrostatic and transport modeling of 2D Topological insulator field-effect transistor (2D TIFETs), based solely on first-principles calculations using density functional theory (DFT). We find that careful consideration of basis set and symmetry constraints in DFT calculations is crucial for determining critical electric field ($E_c$), defined as the electric field intensity at which the topological phase transition occurs. Using ballistic Landauer-Büttiker formula and local potential profile, the drain current-gate bias voltage ($I_D$ - $V_G$) characteristics were obtained and switching behavior was studied. A comparison with the $\mathbf{k} \cdot \mathbf{p}$ model reveals the necessity of DFT calculations for investigating realistic edge dispersions. Our approach provides an efficient and rigorous simulation methodology for mesoscopic transport in 2D TIFETs.

**Keywords**: Topological insulator, quantum spin Hall insulator, first principles calculation, density functional theory, quantum transport theory, Landauer-Büttiker formula

# I. INTRODUCTION

2D topological insulators (2D TIs), also called quantum spin Hall insulators (QSHIs), are quantum materials that exhibit the quantum spin Hall effect, characterized by an insulating bulk and topologically protected helical edge states.[1–5] Within these edge states, backscattering is fundamentally prohibited, leading to the formation of dissipationless conducting channels along the edges. This unique property arises from band inversion and spin-orbit coupling (SOC), and is protected by time-reversal symmetry. When a sufficiently strong perpendicular electric field, known as the critical electric field ($E_c$), is applied, it breaks inversion symmetry and opens band gap ($E_g$). As a result, the material undergoes a phase transition from a topological insulator (TI) to a normal insulator (NI), leading to the vanishing of the conducting edge states[6–9]. This electrically driven phase transition enables the implementation of 2D topological field-effect transistors (2D TIFETs)[10,11].

Owing to their dissipationless edge conduction and electrical tunability, 2D TIFETs are promising candidates for future low-power electronic devices. However, while the TI–NI transition is well-understood physical phenomenon, there remains a critical need for a unified simulation framework that can rigorously and efficiently bridge first-principles electrostatics with mesoscopic transport modeling at the device level. Developing such a methodology is essential for accurately predicting the performance of realistic topological devices.

Since the novel properties of 2D TIs strongly depend on symmetry effects, an appropriate simulation method is essential for their investigation. In particular, the presence of magnetic impurities at the edge or surface of a material breaks the time-reversal symmetry and can significantly alter its topological properties[7]. Moreover, since the edge reconstruction can make edge magnetism, it is important to consider whether the method can account for realistic edge configurations[12]. The **k · p** model and tight-binding (TB) model, owing to its intuitive model Hamiltonian and relatively low computational cost, are frequently used to investigate the electronic properties of large systems, including 2D TI nanoribbons[4–6,8,11,13–18]. However, it not only fails to fully account for realistic edge configurations, but also oversimplifies the interactions within the system that must be considered for rigorous calculations. First-principles calculations based on density functional theory (DFT)[19,20] fully account for all interactions and edge configurations of the system that it provides an accurate description of the microscopic electronic structure without empirical parameters. In addition, it precisely calculates the charge distribution under the presence of perpendicular electric field ($E_z$). Therefore, the symmetry breaking and the topological phase transition can be studied rigorously while previous works often did not consider the symmetry constraints during DFT calculations[6,8,15].

Although the non-equilibrium green function (NEGF) method[21,22] captures quantum transport properties incorporating scattering events, it is computationally too expensive for routinely-conducted calculations or full device modeling. Since the 2D TIs exhibit dissipationless edge transport, meaning there are no scattering events, the transport modeling can be formulated in the ballistic transport regime. In this limit, the NEGF method reduces to the simple ballistic Landauer-Büttiker formalism[23,24]. The top-of-the-barrier (ToB) model utilizes the Landauer formula for barrier-controlled devices while neglecting the finite channel length effects[25]. This approach has been widely used to investigate the performance of nanotransistors in the ballistic trasnport regime, such as nanowire FETs[26,27]. However, the 2D TIFETs are not barrier-controlled devices, the standard ToB model is not directly applicable. Adapting the core logic of the ToB model, wherein the finite channel length effects are neglected, we modified its specific implementation to account for switching mechanism of 2D TIFETs—electrically driven topological phase transition. Thus, first-principles DFT calculation combined with ballistic Landauer-Büttiker formula successfully reflects electrostatics and dissipationless transport characteristics of 2D TIFETs.

In this paper, we develop a first-principles DFT simulation framework for electrostatics and transport modeling of 2D TIFETs with well-known 2D TI, monolayer 1T′-MoS$_2$, as the channel material[8]. We reveal the significance of symmetry constraints during DFT calculation in the aspect of topological phase transition. Using ballistic Landauer-Büttiker formula and electrostatics, we obtain $I_\mathrm{D}$-$V_\mathrm{G}$ characteristics and analyze the switching properties of 2D TIFETs. We highlight the necessity and feasibility of DFT calculation for realistic edge configurations by directly comparing the electrostatics and transport properties of $\mathbf{k}\cdot\mathbf{p}$ model and DFT calculations. Our model can be broadly applied to various 2D TI materials.

# II. ELECTRONIC STRUCTURE CALCULATIONS

First-principles calculations based on DFT[19,20] were conducted using the pseudo-atomic orbital (PAO) based software package OpenMX[28] with 220 Ry of real-space grid cutoff energy and the plane wave (PW) based software package VASP[29–33] with 500 eV of plane-wave cutoff energy. For both codes, the exchange-correlation functional was described within the generalized gradient approximation (GGA) as formulated by Perdew, Burke, and Ernzerhof (PBE)[34]. The self-consistent field (SCF) loop convergence criterion was set to $10^{-6}$ eV for both codes. 9×9×1 and 9×1×1 meshes were used for bulk and nanoribbon calculations, respectively, to sample the first Brillouin zone. Non-collinear spin texture and SOC effects were considered in all calculations. To eliminate spurious interactions between monolayers, a vacuum region of 15 Å was added along the z-direction for both bulk and nanoribbon calculations. For atomic position relaxation, the force tolerance was set to 0.015 eV/Å for OpenMX and 0.01 eV/Å for VASP. The atomic positions and cell volume of the bulk unit cell were optimized while keeping the cell shape fixed. The nanoribbon structures were constructed directly from the optimized bulk unit cell and were not further relaxed to ensure the study remains focused on the resulting electronic structure rather than edge reconstruction effects. A uniform electric field was applied using a sawtooth-like potential, as implemented in both the VASP and OpenMX packages.

We used the continuum $\mathbf{k} \cdot \mathbf{p}$ Hamiltonian and parameters for 1T′-MoS$_2$ from Das et al.[6], with a parameter $\alpha$ updated to reproduce the symmetry constraints considered $E_c$. The Hamiltonian was discretized into a tight-binding-like form using Kwant[35]. In the discretization, the nanoribbon length, width, and grid spacing were set to 2 Å, 300 Å, and 2 Å, respectively. After discretization, the nanoribbon band structure was computed using the same tool.

## III. GEOMETRIES

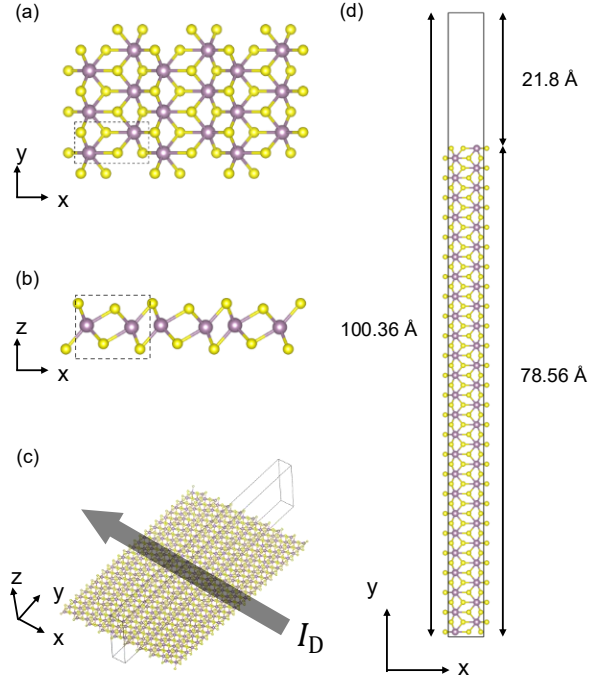

Fig. 1 Geometric structure of bulk and nanoribbon 1T′-MoS$_2$. (a) top view and (b) side view of bulk 3x3 1T′-MoS$_2$ supercell, (c) 3D view and (d) top view of monolayer 1T′-MoS$_2$ nanoribbon. Dashed line in (a), (b) and solid line in (c) denotes unit cell boundary of bulk and nanoribbon, respectively. The direction of the $I_D$ is depicted as transparent gray arrow in (c). Gray spheres represent Mo atoms and yellow sphere represent S atoms.

Monolayer MoS$_2$ can have different structural phases such as stable trigonal prismatic (1H) phase and meta stable octahedral (1T) phase. Due to the instability of 1T phase, it experiences spontaneous structural phase transition into 1T′ phase[36–38]. While 1H phase is normal semiconductor and 1T phase is metal, monolayer 1T′-MoS$_2$ is well known 2D TI with 45 meV of band gap ($E_g$)[8]. As shown in Fig. 1(a, b), 1T′-MoS$_2$ has rectangular primitive cell with inversion symmetry and can be described by space group, $P2_1/m$[39].

Owing to its rectangular unit cell, nanoribbon can easily be made by repeating the unit cell along a direction of a vacuum region to be added. In this work, unit cell is repeated 25 times along the y-direction because its band structure shows helical edge states more explicitly than a vacuum region along x-direction[7]. Then, 21.8 Å of vacuum region was added. The number of repetitions and size of vacuum region were determined to avoid unintentional interactions between edge atoms while considering computational cost. Fig. 1(d) shows top view of nanoribbon unit cell used as channel of device, having bulk-edge-vacuum configuration along y-direction. 3D view of supercell of nanoribbon in Fig. 1(c) shows bulk-edge-vacuum configuration more explicitly Although Figs. 1(c, d) are not depicting actual device geometry, the nanoribbon infinitely long and periodic along the *x*-direction allows for the calculation of the 1D band structure. Although the actual device has a finite channel length, its effects are neglected here, consistent with the ToB approach[25]. From the fact that 2D TI's dissipationless conducting channel exists along the edge between its bulk and surrounding vacuum region, we define x-direction as the direction of drain current ($I_D$) and drain bias voltage ($V_D$). Placing nanoribbon in the x-y plane, introducing $E_z$ can be understood as locating gate electrodes along z-direction.

# IV. ELECTROSTATICS MODELING

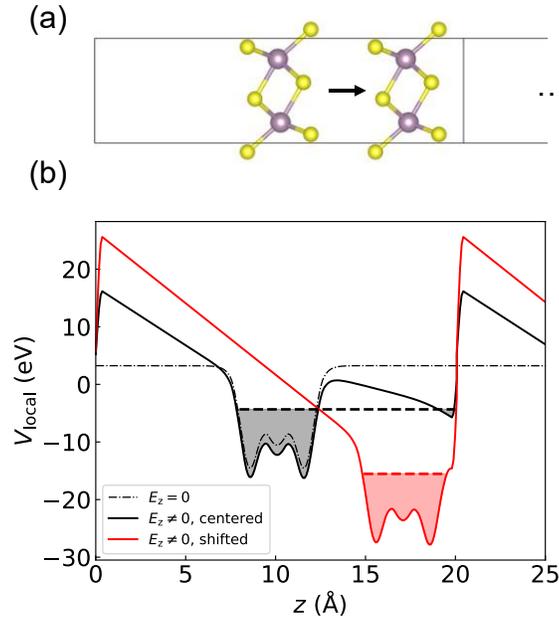

Fig. 2 (a) Side view of centered and shifted bulk 1T′-MoS$_2$. The black arrow depicts the shifting of the atoms. (b) Local potential of centered (black solid line) and shifted (red solid line) bulk 1T′-MoS$_2$. Local potential was averaged over x-y plane to see z dependence of local potential. Region where z > 20 Å is plotted by repeating the local potential periodically, abrupt increase of local potential due to sawtooth-like potential and periodic boundary condition (PBC) is shown at the border of the unit cell. Dashed lines denote Fermi level and dash-dotted line represent unbiased ($E_z = 0$) potential profile of centered 1T′-MoS$_2$.

To simulate topological phase transition and electrostatics of 2D TIs, careful consideration of basis set selection is crucial. Specifically for PW based calculations, the 'electron spilling' problem has been reported in other studies[40,41]. Electron spilling occurs when a large perpendicular electric field is applied along z-direction. As shown by the shaded region in Fig. 2(b), electrons can exist in regions where the energy is between the Fermi level ($E_F$) and the local potential. Therefore, in principle, the vacuum region should not be shaded. However, at $z = 19.5$ Å, a shaded region appears in the vacuum because the uniform electric field, implemented as sawtooth-like potential, lowers the local potential below the $E_F$. This leads to the formation of a spurious virtual state and obstructs a valid simulation of the topological phase transition, making it impossible to determine the $E_c$. The electron spilling problem can be solved by shifting the 2D material along z-axis. As shown in the red curve of Fig. 2, uniform electric field lowers local potential, but no longer below the $E_F$. Therefore, electron does not spill into vacuum region and $E_c$ can be successfully found.

Table 1 Comparison of $E_c$ with and without symmetry constraints.

| Materials | Calculated $E_c$ (V/Å) | | Other works | | |
| --- | --- | --- | --- | --- | --- |
| | With symmetry constraints | Without symmetry constraints | $E_c$ (V/Å) | Calculation Tool | Reference |
| 1T′-MoS$_2$ | 0.07 | 1.3 | 0.142 | VASP | [8] |
| | | | 0.17 | QuantumATK | [6] |
| | | | 1.7 | QuantumATK | [7] |
| MoGe$_2$P$_4$ | 0.08 | 5.0 | 0.077 | VASP | [42] |
| Germanene | < 0.03 | 0.27 | 0.25 | VASP | [43] |
| Stanene | 0.1 | 0.7 | 0.68 | VASP | [15] |

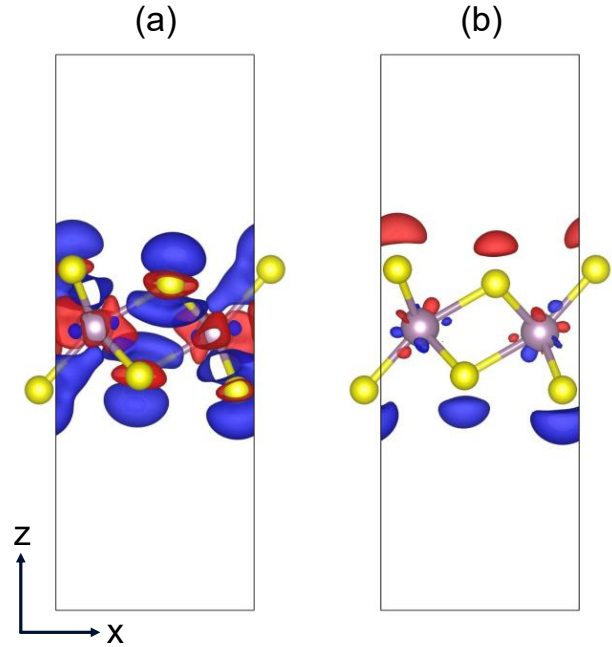

Fig. 3 Charge density difference (Δ$\rho$) of with (a) and without (b) symmetry constraints. The charge density difference was obtained by subtracting the charge density of the system without an electric field from that with an electric field. Isosurfaces are plotted with a value of 0.001 e/Å$^{-3}$, with positive (negative) values colored red (blue).

Symmetry constraints are another critical factor in determining the $E_c$. We calculated $E_c$ for various 2D TIs with and without symmetry constraints (Table 1). A significant difference between the two cases is found and our $E_c$ values also differ from some of those reported in other works. We hypothesize that the large discrepancy between our results and those from other works stems from the use of symmetry constraints in their calculations, given that their reported values are very similar to our values obtained with symmetry constraints.

The difference between with and without symmetry constraints can be explained by plotting the charge density difference (Δ$\rho$). When symmetry constraints are turned off, the inversion symmetry of the charge density is allowed to break naturally, and the resulting Δ$\rho$ (Fig. 3(b)) clearly shows this through regions of equal magnitude but opposite sign at opposing positions. In contrast, when symmetry constraints are imposed, the charge density is forced to maintain inversion symmetry, and the resulting Δρ (Fig. 3(a)) exhibits regions of equal magnitude and same sign at opposing positions. Consequently, to accurately investigate the topological phase transition and determine $E_c$, symmetry constraints should be switched off in the calculations.

The issues of electron spilling and symmetry constraints, are not found in PAO-based calculations. Furthermore, the atomic orbital basis set is known to be more efficient for large systems and for those containing significant vacuum regions[44]. Since our goal is to establish a robust simulation framework for 2D TIFETs, the chosen basis set must be advantageous for modeling systems under a large electric field and with a vacuum region. Therefore, we use PAO-based DFT calculations throughout this work for our first-principles simulations.

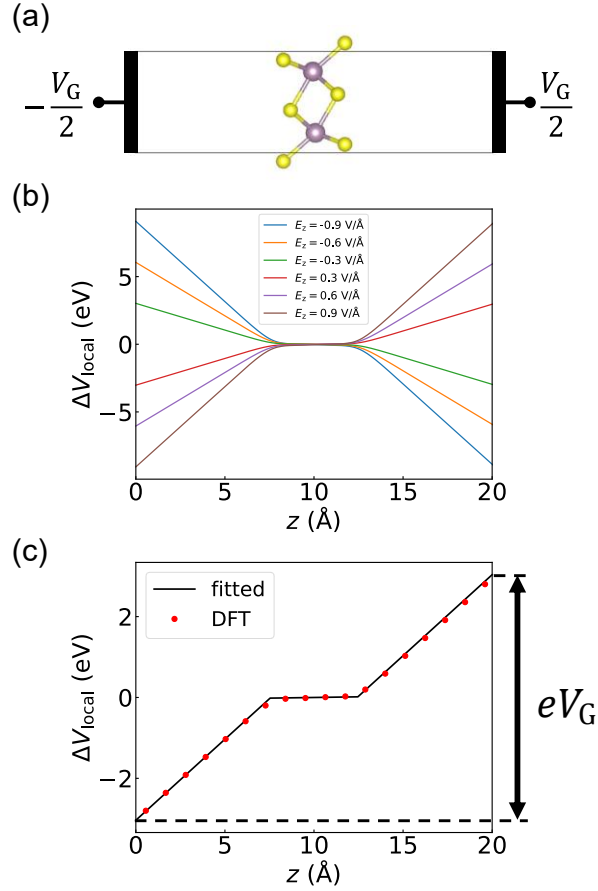

Fig. 4 (a) Schematic diagram of gate electrodes and channel material, where the dielectric is modeled as vacuum. (b) Local potential deference along z-direction of monolayer 1T′-MoS$_2$, as $E_z$ is varied from $-0.9$ V/Å to $+0.9$ V/Å. (c) Simplified local potential deference fitted to DFT result when $E_z = +0.3$ V/Å. For visual clarity, DFT result was plotted as red dot while the fitted curve was plotted as black solid line.

As discussed in Section III, introducing $E_z$ is equivalent to locating gate ~~bias~~ electrodes along z-direction (Fig. 4(a)) and the gate bias voltage ($V_G$) can be acquired by interpreting local potential difference ($\Delta V_{local}$). The top and bottom gates (right and left electrodes) are biased antisymmetrically at $\pm V_G/2$. This enables us to assume that the electrochemical potential of channel material is pinned at 0 eV. Therefore, the net induced carrier density remains at charge neutrality point and the field screening effects can be neglected. The $\Delta V_{local}$ shown in Fig. 4(b, c) is defined as the difference in $V_{local}$ with and without $E_z$, illustrating the influence of $E_z$ on the $V_{local}$. When $E_z = -0.9$ V/Å, $\Delta V_{local}$ drops linearly in the vacuum region. However, near $z = 10$ Å, where channel material is located, $\Delta V_{local}$ drops linearly but almost flat due to the screening effect. This tendency holds regardless of the direction and intensity of $E_z$, as shown in Fig. 4(b).

The ratio of potential slopes in the vacuum and channel material corresponds to the dielectric constant ($\varepsilon_r$). Furthermore, the distance between two points where the potential slope significantly changes can be interpreted as the effective thickness ($t_{eff}$) of the channel material. In this way, we approximated curvilinear $\Delta V_{local}$ into simplified linear $\Delta V_{local}$, as illustrated in Fig. 4(c). $\varepsilon_r$ and $t_{eff}$ were set to fit with DFT result, 44.56 and 4.81 Å were used respectively. Fig. 4 shows the fictitious situation where the vacuum plays the role of dielectric. Thus, one can imagine the situation where the simulation size of the vacuum region is increased, and the $V_G$ is correspondingly increased. Using basic electrodynamics, $V_G$ can be calculated as

$$V_G = \frac{E_z}{\varepsilon_r} \times t_{eff} + \frac{E_z}{3.9} \times \text{EOT}. \tag{1}$$

The equivalent oxide thickness (EOT) was introduced to describe general situation for any dielectric material. The EOT can be replaced flexibly with values specific to the dielectric material and thickness used in the transport modeling. While the incomplete screening effects in single-gated MoS2 FETs have been reported[45], our work utilizes a dual-gated device structure that effectively suppresses these effects.

# V. TRANSPORT MODELING

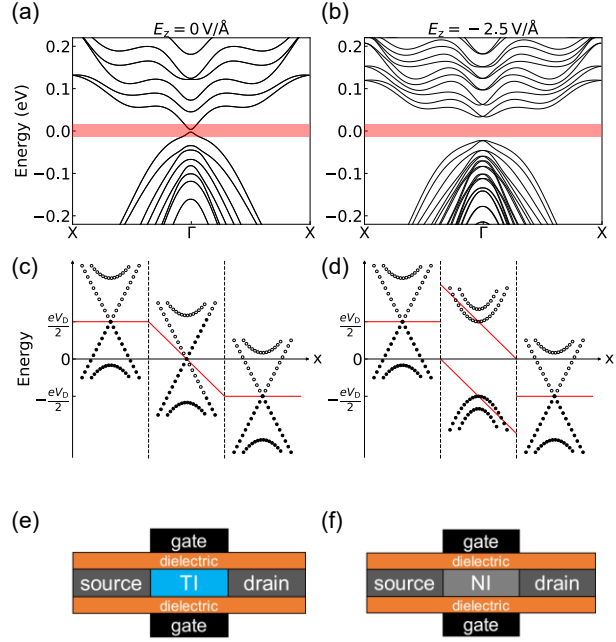

Fig. 5 Band structure of monolayer 1T′-MoS$_2$ (a) with $E_z = 0$ V/Å, (b) with $E_z = -2.5$ V/Å. The red box shows the range of $eV_D = 0.03$ eV. (c, d) Schematic band diagram in real space, schematic band structure is superimposed onto $E_F$ and electric potential energy (solid red lines). Empty circles denote empty states and filled circles denote filled states. (e, f) Device geometries of the dual-gate 2D TIFETs. (a, c, e) and (b, d, f) reflect on- and off-states, respectively.

2D TI's edge conducting channel can be theoretically interpreted by analysing nanoribbon band structure. Fig. 5(a) shows band structure of monolayer 1T′-MoS$_2$ nanoribbon without $E_z$. Electronic states near Γ point of valence band and conduction band are referred as edge states while there still tiny band gap exists due to the finite size effect[17].

The transport modeling is inspired by the ToB model[25]. However, the ToB model is based on barrier-controlled devices, it cannot be directly applied to 2D TIFETs. Fig. 5 (c) depicts schematic band diagram of 2D TIFETs. Instead of potential barrier assumed in ToB model, our model assumes linearly decreasing electrochemical potential. It should be noted that Fig. 5(c) depicts a continuous transition of the electrochemical potential across the channel for illustrative purposes. But, in the ballistic transport regime, the potential drop occurs near the contacts rather than uniformly across the channel. This conceptual visualization serves to clarify the state-filling mechanism and the resulting 'on-state' current. The injection velocity is considered as the group velocity calculated from nanoribbon band structure (channel region).

In the absence of $E_z$ (Fig. 5(a)), monolayer 1T′-MoS$_2$ is in TI phase and retains edge states. By treating the channel as electrically grounded via the source reservoir, the net current is generated by applying $V_D$ in the TI phase, which we denote as the 'on-state'. Fig. 5(c, e) present a conceptual schematic of the band diagram and the dual-gate device geometry, respectively, to facilitate a qualitative understanding of the switching mechanism. In Fig. 5(c), applied drain bias voltage $V_D$ makes Fermi level shift, $eV_D/2$ up in the source region ($E_{FS} = E_F + eV_D/2$) and $eV_D/2$ down in the drain region ($E_{FD} = E_F - eV_D/2$). Thus, electronic states in source (drain) region will be filled up to $E_{FS}$ ($E_{FD}$) by definition. In contrast, in the channel region, electronic states which have positive group velocity ($v_{nk}$) will be filled up to $E_{FS}$ and electronic states which have negative $v_{nk}$ will be filled up to $E_{FD}$. Since $E_{FS}$ is higher than $E_{FD}$, the number of electronic states with positive $v_{nk}$ is larger than the ones with negative $v_{nk}$. As a result, net current will flow from drain to source. The exact amount of drain current ($I_D$) can be calculated by ballistic Landauer-Büttiker formula expressed as

$$I_D = \sum_n \frac{e}{2\pi} \int_{BZ} f_{nk} v_{nk} dk = \sum_n \frac{e}{h} \int_{BZ} f_{nk} \frac{dE_{nk}}{dk} dk. \tag{2}$$

Here, $n$ is band index, $e$ is the electron charge and $h$ is Plank's constant. $E_{nk}$ is the energy dispersion for band $n$ with wave vector $k$, and $f_{nk}$ is Fermi-Dirac distribution function,

$$f_{nk} = \begin{cases} \dfrac{1}{1 + e^{(E_{nk}-E_{\text{FS}})/k_\text{B}T}}, & v_{nk} > 0 \\ \dfrac{1}{1 + e^{(E_{nk}-E_{\text{FD}})/k_\text{B}T}}, & v_{nk} < 0 \end{cases} \tag{3}$$

where $k_\text{B}$ is Boltzmann's constant and $T$ is temperature. As we conducted non-collinear spin formalism during DFT calculations, a factor of two for spin degeneracy is not included in equation (2), but all electronic states are explicitly accounted for by the summation over band index, $n$. Owing to the fact that the transport in 2D TIs is confined to discrete 1D edge channels, the Landauer-Büttiker formula directly yields the total current ($I_\text{D}$) rather than density. As the Fermi-Dirac distribution function depends on the $E_{\text{FS}}(E_{\text{FD}})$, $v_{nk}$, and $T$, equation (2) accounts for the effects of both temperature and drain bias.

The 'off-state' can be achieved by applying $E_\text{z}$. While $E_\text{F}$ and occupation in source and drain region are same as the 'on-state', $E_\text{z}$ opens $E_\text{g}$ and splits valence band maximum (VBM) and conduction band minimum (CBM) in the channel region, as shown in Fig. 5(b) and Fig. 5(d). When $E_\text{z}$ is small, CBM and VBM are still within the range of $\pm eV_\text{D}/2$ but the number of occupied electronic states will be reduced, leading to a decrease in $I_\text{D}$. When $E_\text{z}$ is sufficiently large that CBM and VBM are higher(lower) than $\pm eV_\text{D}/2$, the number of electronic states with positive and negative $v_{nk}$ are same, thus no net current will be generated, i.e., the off-state.

# VI. RESULTS

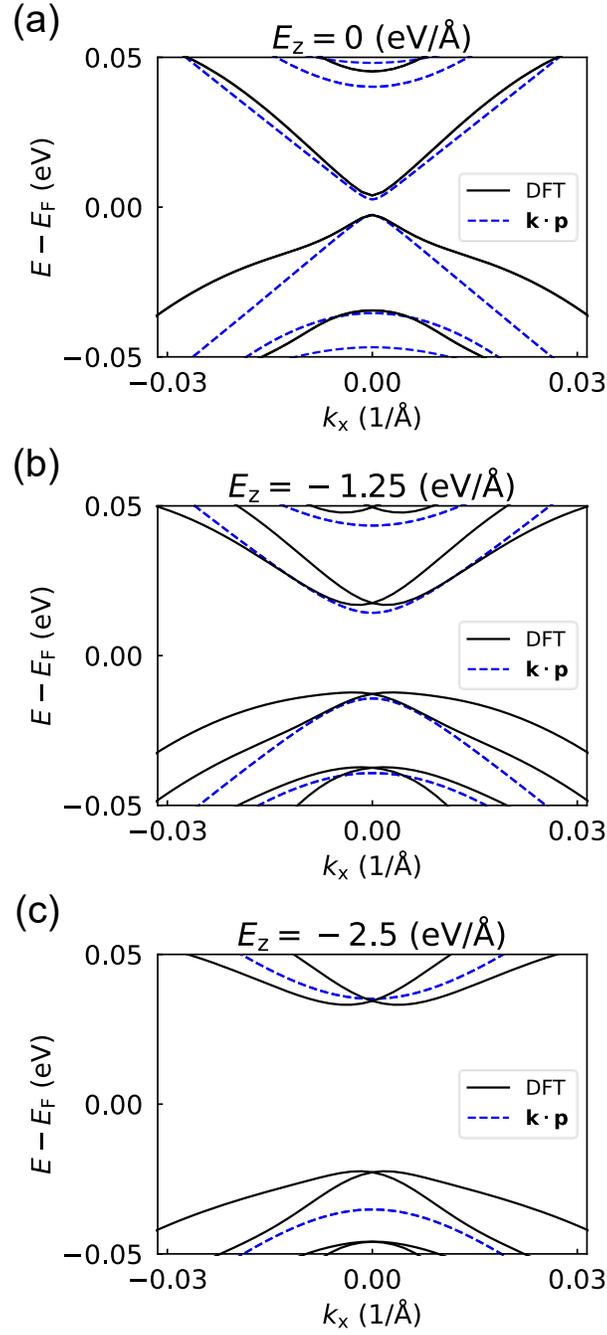

Fig. 6 Nanoribbon band structures calculated using DFT (black solid line) and $\mathbf{k} \cdot \mathbf{p}$ model (blue dashed line) for (a) $E_z = 0$ V/Å ($V_G = 0$ V), (b) $E_z = -1.25$ V/Å ($V_G = -3.2$ V), (c) $E_z = -2.5$ V/Å ($V_G = -6.4$ V).

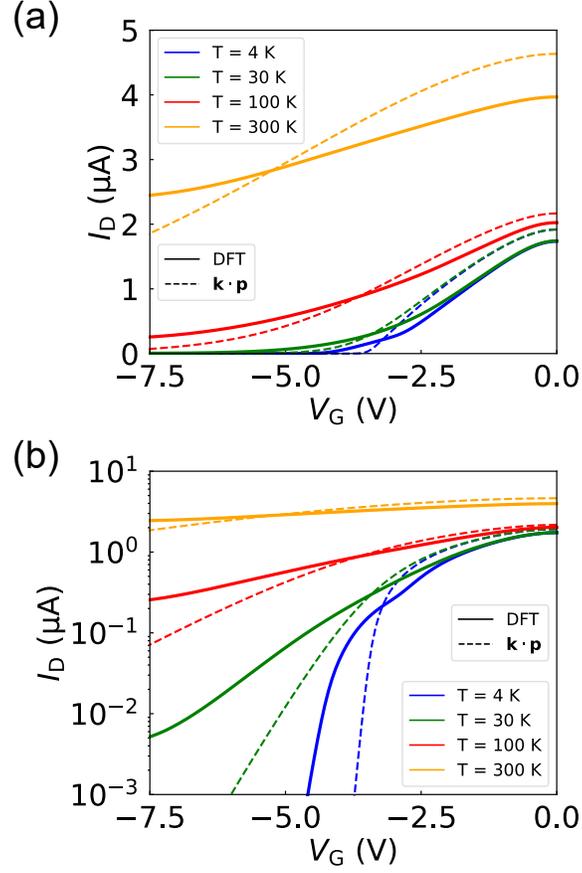

Fig. 7 Calculated $I_D$-$V_G$ characteristics using DFT (solid lines) and $\mathbf{k} \cdot \mathbf{p}$ model (dashed lines) at various temperatures, shown in (a) linear scale and (b) semi-logarithmic scale. $E_z$ were varied from 0 to $-2.85$ V/Å (corresponding to a $V_G$ from 0 to $-7.55$ V) and the temperature from 4 K to 300 K. The necessary parameters for transport modeling were used including $V_D$ of 0.03 V, EOT of 10 nm.

For the DFT results, the on-current ($I_{on}$), which is defined as the drain current $I_D$ at $V_G = 0$ V, was obtained as 1.737, 1.742, 2.025, and 3.969 µA at $T = 4, 30, 100,$ and 300 K, respectively. The off-voltage ($V_{off}$), which is defined as the $V_G$ where $I_D$ becomes $I_{off}$, two orders of magnitude lower than the $I_{on}$, was obtained as $-4.226$ V and $-6.157$ V at 4 K and 30 K, respectively. But at 100 K and 300 K, it does not reach to $I_{off}$ until $V_G = -7.5$ V. Regarding that the modern CMOS technology operates below 1 V, we note that monolayer 1T′-MoS$_2$ channel have significantly large range of operational voltage ranges. The high operational voltage ranges originate from poor topological phase tunability of 1T′-MoS$_2$ via electric field. It has been reported that weak spin-orbit coupling (SOC) strength and strong Rashba splitting strength can enhance topological phase tunability[11,17]. In addition, recent studies found that introducing negative capacitance (NC)—replacing dielectrics by ferroelectrics—can enhance switching properties of 2D TIFETs[46,47]. Thus, further discussions can be made to find alternative 2D TIs materials exhibiting weak SOC and strong Rashba splitting, while incorporating NC into device modeling.

The $I_{on}$ and the $V_{off}$ differ due to the temperature dependence accounted for in the Fermi-Dirac distribution function. As the temperature increases, the Fermi-Dirac distribution broadens, leading to more electronic states with positive group velocity being occupied. This results in an increase in the $I_{on}$. Same analogy applies to $V_{off}$, at high temperatures, more current flows, making it more difficult to turn off the device. Similarly, in Fig. 7(a), a drastic change in the slope of $I_D$ is observed for the $T = 4$ K (blue curve) in the range of $V_G = -5$ to $-2.5$ V. This is attributed to the step-function-like Fermi-Dirac distribution at low temperatures, which causes the number of bands contributing to $I_D$ to change abruptly as the $E_z$ opens the band gap. In contrast, for the cases where $T$ is higher than 30 K, the slope changes smoothly, as the Fermi-Dirac distribution function is no longer step-function-like but is instead thermally broadened.

For the $\mathbf{k} \cdot \mathbf{p}$ model results, the $I_{on}$ was obtained as 1.917, 1.920, 2.167, and 4.635 µA at $T = 4$, 30, 100, and 300 K, respectively. These values are slightly large values than those obtained from the DFT calculations. The $V_{off}$ was obtained as $-3.521$ V and $-4.810$ V at 4 K and 30 K, respectively. Consistent with the DFT results, the $I_D$ does not reach to $I_{off}$ at 100 K and 300 K, until $V_G = -7.5$ V. In contrast to the $I_{on}$, the $\mathbf{k} \cdot \mathbf{p}$ model yields less $V_{off}$ compared to the DFT results.

These differences stem from oversimplified Hamiltonian of the **k · p** model, which is derived using the effective mass approximation and only accurately fits the DFT band structure near $k_x = 0$. As shown in Fig. 6(a), while the band dispersions of the two models are very similar near $k_x = 0$, they begin to differ significantly as $k_x$ is moving away from this point. Specifically, the valence band of **k · p** model has a steeper dispersion than the DFT result. This means electronic states contributing to $I_D$ have a greater group velocity which explains why the $I_{on}$ of the **k · p** model is larger than that of the DFT result. A similar explanation applies to the $V_{off}$. In Fig. 6(b), when a small $E_z$ is applied, the band gap opened by $E_z$ is similar for both models. However, at a large $E_z$, the band gap of the **k · p** model becomes larger than that of the DFT result, as seen in Fig. 6(c). Furthermore, the DFT band structure exhibits horizontally split dispersion in the presence of $E_z$, while the **k · p** model does not. Noting that realistic edge configuration can also alter the dispersion of edge states and that play a crucial role in the $I_D$-$V_G$ characteristics, using DFT to obtain nanoribbon band structure is essential for precise modeling of 2D TIFETs.

## VII. CONCLUSION

We have developed a comprehensive electrostatics and transport modeling framework for 2D TIFETs based solely on first-principles DFT calculations using atomic orbital basis set, which is particularly advantageous for 2D nanoribbon calculations. For the channel material, we used a monolayer 1T′-MoS$_2$ nanoribbon, carefully selecting its size and orientation to reflect the novel properties of a 2D TI. We used a simplified local potential difference to extract parameters from the 2D material, which were then used to convert the $E_z$ into $V_G$. The ballistic Landauer-Büttiker formula was employed to calculate the drain current, incorporating the effects of temperature and drain bias through the Fermi-Dirac distribution function. Using this approach, $I_D$ - $V_G$ characteristics were obtained and switching properties were demonstrated. A comparison with the $\mathbf{k} \cdot \mathbf{p}$ model revealed that DFT is essential for rigorous device modeling, as it is capable of accounting for realistic edge configurations.

**Author contributions**

H. C.: investigation, methodology, software, resources, data curation, formal analysis, visualization and writing – original draft. Y. P.: software, resources, writing – review & editing. S. L.: software, resources, writing – review & editing. Y. L.: conceptualization, methodology, formal analysis, validation, writing – review & editing, resources, funding acquisition, supervision and project administration

**Conflicts of interests**

There are no conflicts to declare.

## Data availability

The data supporting this article have been included as part of the Supplementary Information.

## Acknowledgment

This work was supported by Incheon National University Research Grant in 2024.

**Supplementary Information**

MoS2_NR.cif

```
#======================================================================
# CRYSTAL DATA
#----------------------------------------------------------------------
data_VESTA_phase_1

   _chemical_name_common                 '   MoS2  '
   _cell_length_a                        5.724605
   _cell_length_b                        100.355682
   _cell_length_c                        20.166178
   _cell_angle_alpha                     90.000000
   _cell_angle_beta                      90.000000
   _cell_angle_gamma                     90.000000
   _cell_volume                          11585.400585
   _space_group_name_H-M_alt             'P 1'
   _space_group_IT_number                1

loop_
_space_group_symop_operation_xyz
   'x, y, z'

loop_
   _atom_site_label
   _atom_site_occupancy
   _atom_site_fract_x
   _atom_site_fract_y
   _atom_site_fract_z
   _atom_site_adp_type
```

```
_atom_site_U_iso_or_equiv
_atom_site_type_symbol
Mo1      1.0    0.802396    0.023722    0.496353    Uiso  ?  Mo
Mo2      1.0    0.802396    0.055352    0.496353    Uiso  ?  Mo
Mo3      1.0    0.802396    0.086982    0.496353    Uiso  ?  Mo
Mo4      1.0    0.802396    0.118612    0.496353    Uiso  ?  Mo
Mo5      1.0    0.802396    0.150241    0.496353    Uiso  ?  Mo
Mo6      1.0    0.802396    0.181871    0.496353    Uiso  ?  Mo
Mo7      1.0    0.802396    0.213501    0.496353    Uiso  ?  Mo
Mo8      1.0    0.802396    0.245131    0.496353    Uiso  ?  Mo
Mo9      1.0    0.802396    0.276760    0.496353    Uiso  ?  Mo
Mo10     1.0    0.802396    0.308390    0.496353    Uiso  ?  Mo
Mo11     1.0    0.802396    0.340020    0.496353    Uiso  ?  Mo
Mo12     1.0    0.802396    0.371650    0.496353    Uiso  ?  Mo
Mo13     1.0    0.802396    0.403280    0.496353    Uiso  ?  Mo
Mo14     1.0    0.802396    0.434909    0.496353    Uiso  ?  Mo
Mo15     1.0    0.802396    0.466539    0.496353    Uiso  ?  Mo
Mo16     1.0    0.802396    0.498169    0.496353    Uiso  ?  Mo
Mo17     1.0    0.802396    0.529799    0.496353    Uiso  ?  Mo
Mo18     1.0    0.802396    0.561428    0.496353    Uiso  ?  Mo
Mo19     1.0    0.802396    0.593058    0.496353    Uiso  ?  Mo
Mo20     1.0    0.802396    0.624688    0.496353    Uiso  ?  Mo
Mo21     1.0    0.802396    0.656318    0.496353    Uiso  ?  Mo
Mo22     1.0    0.802396    0.687948    0.496353    Uiso  ?  Mo
Mo23     1.0    0.802396    0.719577    0.496353    Uiso  ?  Mo
Mo24     1.0    0.802396    0.751207    0.496353    Uiso  ?  Mo
Mo25     1.0    0.802396    0.782837    0.496353    Uiso  ?  Mo
Mo26     1.0    0.197602    0.007907    0.503646    Uiso  ?  Mo
Mo27     1.0    0.197602    0.039537    0.503646    Uiso  ?  Mo
Mo28     1.0    0.197602    0.071167    0.503646    Uiso  ?  Mo
Mo29     1.0    0.197602    0.102797    0.503646    Uiso  ?  Mo
```

| | | | | | | |
|---|---|---|---|---|---|---|
| Mo30 | 1.0 | 0.197602 | 0.134427 | 0.503646 | Uiso | ? Mo |
| Mo31 | 1.0 | 0.197602 | 0.166056 | 0.503646 | Uiso | ? Mo |
| Mo32 | 1.0 | 0.197602 | 0.197686 | 0.503646 | Uiso | ? Mo |
| Mo33 | 1.0 | 0.197602 | 0.229316 | 0.503646 | Uiso | ? Mo |
| Mo34 | 1.0 | 0.197602 | 0.260946 | 0.503646 | Uiso | ? Mo |
| Mo35 | 1.0 | 0.197602 | 0.292575 | 0.503646 | Uiso | ? Mo |
| Mo36 | 1.0 | 0.197602 | 0.324205 | 0.503646 | Uiso | ? Mo |
| Mo37 | 1.0 | 0.197602 | 0.355835 | 0.503646 | Uiso | ? Mo |
| Mo38 | 1.0 | 0.197602 | 0.387465 | 0.503646 | Uiso | ? Mo |
| Mo39 | 1.0 | 0.197602 | 0.419094 | 0.503646 | Uiso | ? Mo |
| Mo40 | 1.0 | 0.197602 | 0.450724 | 0.503646 | Uiso | ? Mo |
| Mo41 | 1.0 | 0.197602 | 0.482354 | 0.503646 | Uiso | ? Mo |
| Mo42 | 1.0 | 0.197602 | 0.513984 | 0.503646 | Uiso | ? Mo |
| Mo43 | 1.0 | 0.197602 | 0.545614 | 0.503646 | Uiso | ? Mo |
| Mo44 | 1.0 | 0.197602 | 0.577243 | 0.503646 | Uiso | ? Mo |
| Mo45 | 1.0 | 0.197602 | 0.608873 | 0.503646 | Uiso | ? Mo |
| Mo46 | 1.0 | 0.197602 | 0.640503 | 0.503646 | Uiso | ? Mo |
| Mo47 | 1.0 | 0.197602 | 0.672133 | 0.503646 | Uiso | ? Mo |
| Mo48 | 1.0 | 0.197602 | 0.703762 | 0.503646 | Uiso | ? Mo |
| Mo49 | 1.0 | 0.197602 | 0.735392 | 0.503646 | Uiso | ? Mo |
| Mo50 | 1.0 | 0.197602 | 0.767022 | 0.503646 | Uiso | ? Mo |
| S1 | 1.0 | 0.083737 | 0.023722 | 0.586074 | Uiso | ? S |
| S2 | 1.0 | 0.083737 | 0.055352 | 0.586074 | Uiso | ? S |
| S3 | 1.0 | 0.083737 | 0.086982 | 0.586074 | Uiso | ? S |
| S4 | 1.0 | 0.083737 | 0.118612 | 0.586074 | Uiso | ? S |
| S5 | 1.0 | 0.083737 | 0.150241 | 0.586074 | Uiso | ? S |
| S6 | 1.0 | 0.083737 | 0.181871 | 0.586074 | Uiso | ? S |
| S7 | 1.0 | 0.083737 | 0.213501 | 0.586074 | Uiso | ? S |
| S8 | 1.0 | 0.083737 | 0.245131 | 0.586074 | Uiso | ? S |
| S9 | 1.0 | 0.083737 | 0.276760 | 0.586074 | Uiso | ? S |
| S10 | 1.0 | 0.083737 | 0.308390 | 0.586074 | Uiso | ? S |

| | | | | | | | |
|---|---|---|---|---|---|---|---|
| S11 | 1.0 | 0.083737 | 0.340020 | 0.586074 | Uiso | ? | S |
| S12 | 1.0 | 0.083737 | 0.371650 | 0.586074 | Uiso | ? | S |
| S13 | 1.0 | 0.083737 | 0.403280 | 0.586074 | Uiso | ? | S |
| S14 | 1.0 | 0.083737 | 0.434909 | 0.586074 | Uiso | ? | S |
| S15 | 1.0 | 0.083737 | 0.466539 | 0.586074 | Uiso | ? | S |
| S16 | 1.0 | 0.083737 | 0.498169 | 0.586074 | Uiso | ? | S |
| S17 | 1.0 | 0.083737 | 0.529799 | 0.586074 | Uiso | ? | S |
| S18 | 1.0 | 0.083737 | 0.561428 | 0.586074 | Uiso | ? | S |
| S19 | 1.0 | 0.083737 | 0.593058 | 0.586074 | Uiso | ? | S |
| S20 | 1.0 | 0.083737 | 0.624688 | 0.586074 | Uiso | ? | S |
| S21 | 1.0 | 0.083737 | 0.656318 | 0.586074 | Uiso | ? | S |
| S22 | 1.0 | 0.083737 | 0.687948 | 0.586074 | Uiso | ? | S |
| S23 | 1.0 | 0.083737 | 0.719577 | 0.586074 | Uiso | ? | S |
| S24 | 1.0 | 0.083737 | 0.751207 | 0.586074 | Uiso | ? | S |
| S25 | 1.0 | 0.083737 | 0.782837 | 0.586074 | Uiso | ? | S |
| S26 | 1.0 | 0.578426 | 0.007907 | 0.565857 | Uiso | ? | S |
| S27 | 1.0 | 0.578426 | 0.039537 | 0.565857 | Uiso | ? | S |
| S28 | 1.0 | 0.578426 | 0.071167 | 0.565857 | Uiso | ? | S |
| S29 | 1.0 | 0.578426 | 0.102797 | 0.565857 | Uiso | ? | S |
| S30 | 1.0 | 0.578426 | 0.134427 | 0.565857 | Uiso | ? | S |
| S31 | 1.0 | 0.578426 | 0.166056 | 0.565857 | Uiso | ? | S |
| S32 | 1.0 | 0.578426 | 0.197686 | 0.565857 | Uiso | ? | S |
| S33 | 1.0 | 0.578426 | 0.229316 | 0.565857 | Uiso | ? | S |
| S34 | 1.0 | 0.578426 | 0.260946 | 0.565857 | Uiso | ? | S |
| S35 | 1.0 | 0.578426 | 0.292575 | 0.565857 | Uiso | ? | S |
| S36 | 1.0 | 0.578426 | 0.324205 | 0.565857 | Uiso | ? | S |
| S37 | 1.0 | 0.578426 | 0.355835 | 0.565857 | Uiso | ? | S |
| S38 | 1.0 | 0.578426 | 0.387465 | 0.565857 | Uiso | ? | S |
| S39 | 1.0 | 0.578426 | 0.419094 | 0.565857 | Uiso | ? | S |
| S40 | 1.0 | 0.578426 | 0.450724 | 0.565857 | Uiso | ? | S |
| S41 | 1.0 | 0.578426 | 0.482354 | 0.565857 | Uiso | ? | S |
| S42 | 1.0 | 0.578426 | 0.513984 | 0.565857 | Uiso | ? | S |
| S43 | 1.0 | 0.578426 | 0.545614 | 0.565857 | Uiso | ? | S |
| S44 | 1.0 | 0.578426 | 0.577243 | 0.565857 | Uiso | ? | S |
| S45 | 1.0 | 0.578426 | 0.608873 | 0.565857 | Uiso | ? | S |
| S46 | 1.0 | 0.578426 | 0.640503 | 0.565857 | Uiso | ? | S |
| S47 | 1.0 | 0.578426 | 0.672133 | 0.565857 | Uiso | ? | S |
| S48 | 1.0 | 0.578426 | 0.703762 | 0.565857 | Uiso | ? | S |
| S49 | 1.0 | 0.578426 | 0.735392 | 0.565857 | Uiso | ? | S |
| S50 | 1.0 | 0.578426 | 0.767022 | 0.565857 | Uiso | ? | S |
| S51 | 1.0 | 0.421576 | 0.023722 | 0.434143 | Uiso | ? | S |
| S52 | 1.0 | 0.421576 | 0.055352 | 0.434143 | Uiso | ? | S |
| S53 | 1.0 | 0.421576 | 0.086982 | 0.434143 | Uiso | ? | S |
| S54 | 1.0 | 0.421576 | 0.118612 | 0.434143 | Uiso | ? | S |
| S55 | 1.0 | 0.421576 | 0.150241 | 0.434143 | Uiso | ? | S |
| S56 | 1.0 | 0.421576 | 0.181871 | 0.434143 | Uiso | ? | S |
| S57 | 1.0 | 0.421576 | 0.213501 | 0.434143 | Uiso | ? | S |
| S58 | 1.0 | 0.421576 | 0.245131 | 0.434143 | Uiso | ? | S |
| S59 | 1.0 | 0.421576 | 0.276760 | 0.434143 | Uiso | ? | S |
| S60 | 1.0 | 0.421576 | 0.308390 | 0.434143 | Uiso | ? | S |
| S61 | 1.0 | 0.421576 | 0.340020 | 0.434143 | Uiso | ? | S |
| S62 | 1.0 | 0.421576 | 0.371650 | 0.434143 | Uiso | ? | S |
| S63 | 1.0 | 0.421576 | 0.403280 | 0.434143 | Uiso | ? | S |
| S64 | 1.0 | 0.421576 | 0.434909 | 0.434143 | Uiso | ? | S |
| S65 | 1.0 | 0.421576 | 0.466539 | 0.434143 | Uiso | ? | S |
| S66 | 1.0 | 0.421576 | 0.498169 | 0.434143 | Uiso | ? | S |
| S67 | 1.0 | 0.421576 | 0.529799 | 0.434143 | Uiso | ? | S |
| S68 | 1.0 | 0.421576 | 0.561428 | 0.434143 | Uiso | ? | S |
| S69 | 1.0 | 0.421576 | 0.593058 | 0.434143 | Uiso | ? | S |
| S70 | 1.0 | 0.421576 | 0.624688 | 0.434143 | Uiso | ? | S |
| S71 | 1.0 | 0.421576 | 0.656318 | 0.434143 | Uiso | ? | S |
| S72 | 1.0 | 0.421576 | 0.687948 | 0.434143 | Uiso | ? | S |
| S73 | 1.0 | 0.421576 | 0.719577 | 0.434143 | Uiso | ? | S |
| S74 | 1.0 | 0.421576 | 0.751207 | 0.434143 | Uiso | ? | S |
| S75 | 1.0 | 0.421576 | 0.782837 | 0.434143 | Uiso | ? | S |
| S76 | 1.0 | 0.916263 | 0.007907 | 0.413927 | Uiso | ? | S |
| S77 | 1.0 | 0.916263 | 0.039537 | 0.413927 | Uiso | ? | S |
| S78 | 1.0 | 0.916263 | 0.071167 | 0.413927 | Uiso | ? | S |
| S79 | 1.0 | 0.916263 | 0.102797 | 0.413927 | Uiso | ? | S |
| S80 | 1.0 | 0.916263 | 0.134427 | 0.413927 | Uiso | ? | S |
| S81 | 1.0 | 0.916263 | 0.166056 | 0.413927 | Uiso | ? | S |
| S82 | 1.0 | 0.916263 | 0.197686 | 0.413927 | Uiso | ? | S |
| S83 | 1.0 | 0.916263 | 0.229316 | 0.413927 | Uiso | ? | S |
| S84 | 1.0 | 0.916263 | 0.260946 | 0.413927 | Uiso | ? | S |
| S85 | 1.0 | 0.916263 | 0.292575 | 0.413927 | Uiso | ? | S |
| S86 | 1.0 | 0.916263 | 0.324205 | 0.413927 | Uiso | ? | S |
| S87 | 1.0 | 0.916263 | 0.355835 | 0.413927 | Uiso | ? | S |
| S88 | 1.0 | 0.916263 | 0.387465 | 0.413927 | Uiso | ? | S |
| S89 | 1.0 | 0.916263 | 0.419094 | 0.413927 | Uiso | ? | S |
| S90 | 1.0 | 0.916263 | 0.450724 | 0.413927 | Uiso | ? | S |
| S91 | 1.0 | 0.916263 | 0.482354 | 0.413927 | Uiso | ? | S |
| S92 | 1.0 | 0.916263 | 0.513984 | 0.413927 | Uiso | ? | S |
| S93 | 1.0 | 0.916263 | 0.545614 | 0.413927 | Uiso | ? | S |
| S94 | 1.0 | 0.916263 | 0.577243 | 0.413927 | Uiso | ? | S |
| S95 | 1.0 | 0.916263 | 0.608873 | 0.413927 | Uiso | ? | S |
| S96 | 1.0 | 0.916263 | 0.640503 | 0.413927 | Uiso | ? | S |
| S97 | 1.0 | 0.916263 | 0.672133 | 0.413927 | Uiso | ? | S |
| S98 | 1.0 | 0.916263 | 0.703762 | 0.413927 | Uiso | ? | S |
| S99 | 1.0 | 0.916263 | 0.735392 | 0.413927 | Uiso | ? | S |
| S100 | 1.0 | 0.916263 | 0.767022 | 0.413927 | Uiso | ? | S |